\documentclass[a4paper,11pt,fleqn]{article}
\usepackage[utf8]{inputenc}
\usepackage{rotating}
\usepackage[T1]{fontenc}
\usepackage{graphicx}
\usepackage{amsmath}
\usepackage{amsfonts}
\usepackage{amssymb}
\usepackage[svgnames]{xcolor}
\usepackage[colorlinks,citecolor=blue,linkcolor=blue]{hyperref}
\usepackage{bm}
\usepackage{float}
\usepackage{multicol}
\usepackage{booktabs}
\usepackage[numbers]{natbib}
\date{}
\definecolor{orcidlogocol}{HTML}{A6CE39}
\usepackage{flexisym}
\usepackage[font={sf},labelfont=bf]{caption}
\definecolor{orRev}{HTML}{D35400}
\definecolor{grRev}{HTML}{008c00}

\title{\textbf{\textsf{On the proposed concept of mechanical phasons\\ in Ni-Mn-Ga modulated martensite}}}
\author{Petr Sedl\'{a}k$^\dag$ {\small \sf ({\color{orcidlogocol}{\textbf{\textsf ORCiD:}}} 0000-0002-0700-902X)}, Tom\'{a}\v{s} Grabec$^\dag$ {\small \sf ({\color{orcidlogocol}{\textbf{\textsf ORCiD:}}} 0000-0002-5211-8217)}, \\ Hanu\v{s} Seiner$^\dag$$^*$ {\small \sf ({\color{orcidlogocol}{\textbf{\textsf ORCiD:}}} 0000-0002-1151-7270)}}

\begin{document}

\maketitle

\begin{center}
\baselineskip18pt
{\small $^\dag$ Institute of Thermomechanics, Czech Academy of Sciences, Dolej\v{s}kova 5, 182 00 Prague, Czech Republic}

$^*$Corresponding author: hseiner@it.cas.cz

\end{center}

\bigskip

\begin{abstract}
We discuss modulation phasons as a possible source of unusual elastic behavior of five-layer modulated (10\,M) martensite of the Ni-Mn-Ga shape memory alloy. This material exhibits anomalous macroscopic shear compliance along specific planes perpendicular to the modulation vector, and this compliance disappears when the modulations become incommensurate. Using a simple mechanical model, we show that modulation phasons in Ni-Mn-Ga can have macroscopic mechanical manifestations, and that the resulting 'mechanical phasons' can relax external shear loadings for commensurate and weakly incommensurate modulations, but not for strongly incommensurate modulations. The model merges ideas from the adaptive martensite theory and electronic-structure considerations, and enables straightforward explanations of several properties of the 10\,M lattice, such as spontaneous monoclinic distortion or easy formation and propagation of $a/b$ twins.
\end{abstract}


\bigskip 

\section{Introduction}

In five-layer modulated (10\,M) martensites of Ni-Mn-Ga-based ferromagnetic shape memory alloys, the stress needed to set twin boundaries into motion falls below 1 MPa \cite{Straka2011,Sozinov2011}, which is a phenomenon often termed twin supermobility \cite{SeinerReview,Heczko2022}. The supermobility makes these alloys prominent candidates for micromechanical applications that require achieving large reversible strains under small mechanical loadings, or in weak magnetic fields \cite{Heczko2022}. The latter functionality follows from the quite strong magnetocrystalline anisotropy of Ni-Mn-Ga, which, together with the low twinning stress, gives rise to the magnetically-induced reorientation effect (MIR, \cite{MIR}), which is also observed only in very few magnetic shape memory alloys \cite{MIR_other1,MIR_other2}. {\color{black}{The low twinning stress, as well as other unusual magneto-mechanical properties of Ni-Mn-Ga, are symptomatic of the five-layer modulations: they are significantly suppressed in seven-layer modulated (14\,M) martensite, and completely disappear in non-modulated (NM) martensite. }} These effects indicate that there is a strong mechanical instability in the 10\,M modulated lattice; the existence of such instability is typically manifested by strong elastic anisotropy resulting from vanishing elastic stiffness against shearing along specific crystallographic planes \cite{Nakanishi}. Such a behavior was confirmed already by the first attempts to determine anisotropic elastic coefficients of Ni-Mn-Ga martensite using ultrasonic methods \cite{Dai}, where specific shear moduli as soft as $c^*\approx 2.5$\,GPa were documented. Later, direct, X-ray diffraction-based stress-strain measurements by Cejpek et al.\cite{Cejpek} have confirmed a low Young's modulus (below 1 GPa) at very small strains, but neither of these approaches enabled linking the elastic instability to specific straining modes of the lattice. This has been done recently by Rep\v{c}ek et al. \cite{Repcek2024}, who revealed that the soft elastic behavior corresponds to shears changing the monoclinic angle in 10\,M lattice $\gamma_{\rm 10M}$. Geometrically, this means that the soft elastic strains are oriented identically as the spontaneous lattice distortions carried by the modulation wave itself. {\color{black}{ The study \cite{Repcek2024} utilized highly accurate laser-ultrasonic experiments, and the measurements were performed on Ni-Mn-Ga crystals with commensurate 10\,M modulations, that is, with the modulation wave period perfectly matching 10 interatomic spacings (see the next section for a more detailed introduction of commensurate/incommensurate modulation functions and main theoretical models of how the modulation arises). For this material, the softest elastic coefficient, denoted $c_{55}$ in the coordinate system used in \cite{Repcek2024},}} was shown to be location-dependent, and reaching as low values as $c_{55}=0.4$\,GPa (and $c_{55}=2.1$\,GPa on average), making Ni-Mn-Ga one of the most elastically anisotropic materials ever reported, and the authors have discussed the possible relation between this softness and supermobility.

At the same time, nevertheless, there are observations that seem to contradict part of the conclusions of \cite{Repcek2024}. Most importantly, Sozinov et al. \cite{Sozinov2026} recently reported on elastic constants of the same crystal lattice (10\,M martensite of a Ni-Mn-Ga-based alloy) but with incommensurate modulations. This material was shown to be much elastically stiffer  with respect to the discussed shearing orientation ($c_{55}=6.4$\,GPa), although it still exhibits twin supermobility. This means that the lattice instability responsible for the low twinning stress does not necessarily manifest macroscopically as very low $c_{55}$. The result of \cite{Sozinov2026} were also in agreement with previous observation by Ve\v{r}t\'{a}t et al.\cite{Vertat}, who reported on pronounced increase in bending stiffness of a plate-like 10\,M Ni-Mn-Ga-Fe single crystal with the commensurate$\rightarrow$incommensurate transition. Secondly, the experimental values both for commensurate and incommensurate 10\,M phases are in a sharp contrast with theoretical predictions based on first-principles calculations, where the reported values for $c_{55}$ range from 15 GPa \cite{OzdemirKart} to 18 GPa \cite{SeinerReview}. This discrepancy is too large to be ascribed to the limited accuracy and reliability of first principles calculations, although the size and complexity of the 10\,M unit cell and the effect of magnetism may put these both under question. Moreover, these calculations are also strongly supported by inelastic neutron scattering (INS) results: Shapiro et al. \cite{Shapiro}  performed INS experiments on incommensurate 10\,M Ni-Mn-Ga crystals, and have determined the slope of the TA$_{2}$ phonon branch in the $[\zeta\zeta0]$ direction to be 1.6 mm.$\mu$s$^{-1}$, which corresponds to $c_{55}=20.5$\,GPa.  The same result was later confirmed independently by Ener et al. \cite{Ener2013}. In other words, at crystal-lattice lengthscales where the INS interaction takes place, the 10\,M lattice is indeed as stiff as predicted by the first-principles calculations.

\begin{figure}[!t]
\vspace{-0.5cm}
 \begin{minipage}[b]{0.55\textwidth}
 \hspace{-1.25cm} \includegraphics[width=1.07\textwidth]{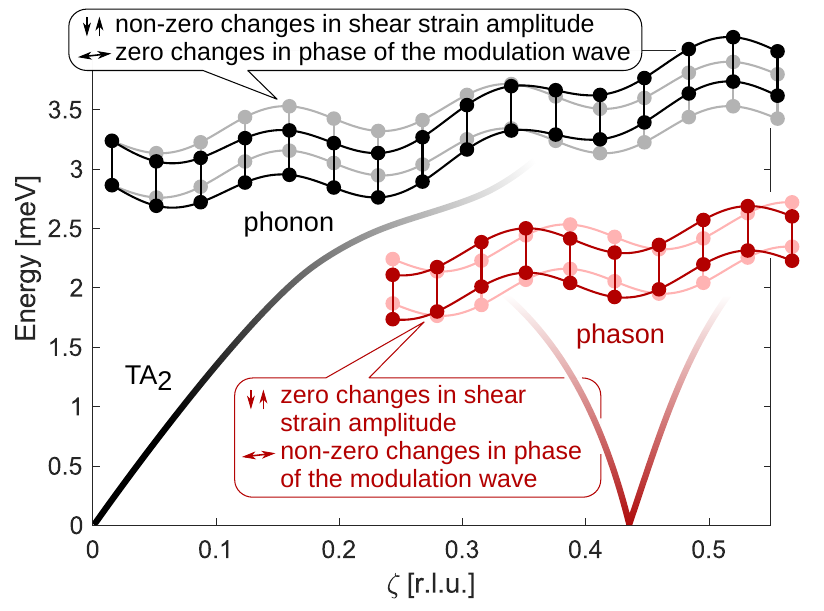}
 \end{minipage}\hfill
\begin{minipage}[b]{0.45\textwidth}
  \caption{Schematic outline of the INS result reported in \cite{Shapiro}: the TA$_2$ branch represents a phonon propagating along the modulation direction and having polarization in the direction of the modulation amplitude, the phason branch represents the same mechanical shearing of the lattice, but through phase shifts in the modulation function.}\label{fig1a}
  \end{minipage}
  
\end{figure}

\begin{figure}[!t]
\vspace{-0.5cm}
 \begin{minipage}[b]{0.55\textwidth}
 \hspace{-1.25cm} \includegraphics[width=1.07\textwidth]{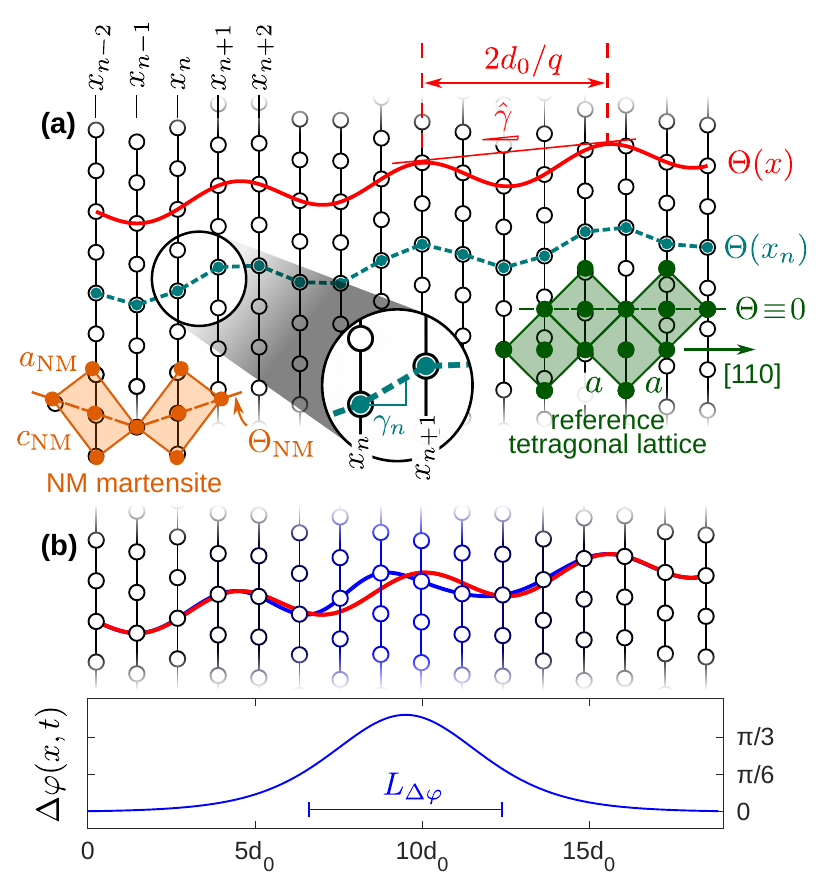}
 \end{minipage}\hfill
\begin{minipage}[b]{0.45\textwidth}
  \caption{Introducing the notation used for the modulation function \textbf{(a)} and for the phason \textbf{(b)}, respectively. For easier visualization, the phason in \textbf{(b)} has an unphysically small length of $L_{\Delta\varphi}\approx{}5d_0$, and an unphysically large amplitude (over $\pi/3$). For the model constructed below, we will assume phasons of lengths of several tens or hundreds of $d_0$ and amplitudes below $\pi/10$. In {(a)}, two close approximations of the moduled lattice using tetragonal lattice are shown: i) the reference tetragonal lattice (with $\Theta\equiv{}0$) from which the modulated lattice  differs just by small displacements carried by the static modulation wave; ii) the nanotwinning construction from non-modulated (NM) building blocks considered in the adaptive concept. For the latter, the modulation function is a sharp zig-zag function $\Theta_{\rm NM}$.}\label{fig1bc}
  \end{minipage}

\end{figure}

\bigskip
In this paper, we discuss a possible explanation of these contradictory results based on mechanics of \emph{modulation phasons} \cite{Phasons}, that is, coherent dynamic shifts of the phase of the modulation function in the 10\,M lattice. We are motivated by the above mentioned INS observations by Shapiro et al. \cite{Shapiro}, who have seen additional phonon-like branches in their INS experiments, with zero-energy points corresponding to the wavelength of modulations (Figure \ref{fig1a}). Shapiro et al. \cite{Shapiro} interpreted this observation as a proof of existence of phasons in the lattice.  Ener et al. \cite{Ener2013} later disagreed on this conclusion, showing very similar INS results but claiming that the branches may originate from the fact that the experiments were performed on a mixture of variants, which may lead to superposing information from $[\zeta\zeta0]$ and $[\zeta\bar{\zeta}0]$ directions and detecting artificial $\Gamma-$points in the INS spectrum. A strong argument in favor of Ener et al. \cite{Ener2013} was that they observed the additional branch in commensurate 10\,M structure, while phasons are typically expected to appear in incommensurate structures, where the modulation wave cannot pin on lattice sites. However, as the above discussed discrepancy in elastic properties reveals there might be additional degrees of freedom in the 10\,M lattice, the idea of phasons in Ni-Mn-Ga needs to be revisited. The relation between phasons and mechanics of the 10\,M lattice has been recently proposed by Ve\v{r}t\'{a}t et al. \cite{Straka2026}, based on their previous detailed analysis of the modulation function of the 10\,M lattice and its evolution with temperature \cite{Vertat2024}. Here we elaborate more on this concept, construct a simple mechanistic model of the phasons, show how the phasons can be responsible for shear softening of the lattice, and how this effect is suppressed in the commensurate$\rightarrow$incommensurate transition. {\color{black} Phasons are supposed to affect the mechanical response in perovskites (e.g. \cite{Bechtle_PRB_1978}), where the modulation wave and elastic strains are coupled through rotation of adjacent octahedra (Fritz's mechanism, \cite{Fritz_PRL_1975}); a mechanical model capturing this behavior based on a double-well Landau energy function was reported in \cite{Poulet_SSP_1981}. We adopt a similar approach in this paper, considering the double-well energy function as a result of meta-stability of the 10\,M lattice with respect to non-modulated martensite.}

Besides the INS experiments \cite{Shapiro,Ener2013}, there is only sparse and indirect evidence of phasons in modulated Ni-Mn-Ga lattices in the literature, such as the peak broadening  in X-ray diffraction reported in \cite{CDW2}. The model presented in this paper can be understood as an additional indirect confirmation of their possible presence. In addition to explaining the unusual elastic behavior observed in laser-ultrasonic experiments, and its dependence on the modulation vector, the proposed concept of phason mechanics could also contribute to rationalizing the stability of various modulated structures in Ni-Mn-Ga and the transformations between them. In particular, the observed temperature-induced transformation between the 14\,M and 10\,M structures has no obvious reason in the adaptive concept. The presence of phasons and the associated low-frequency phonon branches with large vibrational entropy \cite{Phasons} at finite temperatures may explain the stability of the 10M structure at elevated temperatures. In this sense, the model may have broader implications, the discussion of which, however, falls beyond the current scope.

\section{Model construction}

\subsection{Preliminaries}

Structural modulations in Ni-Mn-Ga are slight, periodically repeating displacements of atoms observed as a static modulation wave running through the crystal \cite{Santamarta}. The atoms are displaced with respect to lattice sites of a tetragonal lattice with $c/a<1$ (termed hereafter \emph{reference tetragonal lattice}, see Figure \ref{fig1bc}(a)), in which the modulation wave runs along one of the $\langle{1\,1\,0}\rangle$ directions, and the displacements themselves are perpendicular to this direction and lying in the $(0\,0\,1)$ plane. The resulting modulated lattice than has a {\color{black} monoclinic} symmetry, with lattice parameters very close to those of the reference tetragonal lattice. {\color{black} The wave vector of the modulation wave ${\mathbf q}$ is called the modulation vector. For ${\mathbf n^*}$ being the reciprocal-space vector $\bm{\langle1\,1\,0\rangle}^*$ representing modulation direction, the modulation vector is
\begin{equation}
 {\mathbf q}= q{\mathbf n^*},
\end{equation}
where $q$ is the modulation wave number expressed in reciprocal lattice units (r.l.u)} that in Ni-Mn-Ga  typically attains magnitudes either close to $q=2/5$ (five-layer modulated martensite, 10\,M) or to $q=2/7$ (seven-layer modulated martensite, 14\,M). If the $q$ is exactly equal to one of these values, the modulations are commensurate, if it slightly deviates, the modulations are incommensurate; a specific case is $q=3/7$, which is a close to $q=2/5$ and is referred to as pseudo-commensurate modulations (an orthorhombic structure that can be denoted as 14\,O \cite{Straka2026}).

In the past two decades, several simplified models of modulated lattices in Ni-Mn-Ga-based alloys have been proposed. Among these, the most appraised one was the model based on adaptive modulations \cite{Kaufman_PRL,Kaufman_NJP}, in which the modulation is constructed from blocks of non-modulated (NM) martensite, that is tetragonal with $c_{\rm NM}/a_{\rm NM}>1$ (such that $a_{\rm NM}$ equals to the $c$ parameter of the reference tetragonal lattice). The modulated structure is then understood as a nanotwinned microstructure of two variants of NM martensite differing in the orientations of the $c-$axis. The adaptive concept suggest that the nanotwinning arises due to strain compatibility between martensite and the parent phase, and is stabilized due to energetic proficiency of the $3|\bar{2}|3|\bar{2}$ or $5|\bar{2}|5|\bar{2}$ stacking sequences \cite{Gruner}. The modulation function in the adaptive concept is a zig-zag (triangle) function with altering intervals of constant positive and negative slope, and the possible incommensurateness is achieved only in averaged sense, as a result of stacking faults in the nanotwinning sequence. The adaptive concept gives very good predictions of lattice parameters, especially for 14\,M martensite, and also offers a physically meaningful explanation of gradual coarsening from 14\,M to NM in epitaxial films \cite{Kaufman_NJP}. On the other hand, for 10\,M martensite it overestimates both the modulation amplitude and the monoclinic angle, and cannot directly explain the driving force for the 10\,M$\rightarrow$14\,M intermartensitic transition.

In an alternative approach \cite{CDW1,CDW2}, the modulation is assumed to arise as a result of Fermi-surface nesting, and can be interpreted as a frozen phonon with a specific wave vector, or a charge density wave (CDW). The modulation function in such case is a harmonic function (a sine wave). This approach gives better predictions for lattice parameters of 10\,M martensite, but cannot explain the monoclinic distortion of the modulated lattice, or massive appearance of stacking faults in the modulation sequences \cite{Zarubova}. An attempt to merge both approaches into one model was done in \cite{Benesova}, but the resulting model, while being mathematically rigorous and tractable, did not capture several essential physical features of modulated martensite, such as the monoclinic distortion or spontaneous $a/b$ twinning.

Most recently, Ve\v{r}t\'{a}t et al.\cite{Vertat2024} analyzed extensive sets of X-ray diffraction data to infer details on the modulation function in 10\,M martensite. It turned out that the modulation function cannot be a single sine wave, but has numerous higher-harmonic components, so that it partially adopts the zig-zag character predicted by the adaptive concept. In a continuing research \cite{Straka2026}, it was revealed that the incommensurate structure spontaneously creates interfaces reverting the stacking order, such as $3|\bar{2}|3|\bar{2}||2|\bar{3}|2|\bar{3}$, which are the $a/b$ twins. Even this detailed analysis did not, nevertheless, elucidate the origin of the monoclinic distortion in the 10\,M lattice or the reason why this distortion is so particularly soft with respect to mechanical loading. 

\bigskip
In the following, we take fundamentals of the above approaches, and create a simple mechanistic model of phasons in 10\,M Ni-Mn-Ga, that is, of small {\color{black} perturbations of} the phase of the modulation function. We adopt the constant plane shift (CPS) assumption introduced by Vinogradova et al. \cite{Vinogradova2023}, which means we will consider that the modulated lattice consists of perfectly rigid planes oriented perpendicular to the modulation vector and spaced with fixed distance $d_0$. The rigid planes can shift up and down along the direction of the modulation amplitude perpendicular to the modulation vector. The rigid planes are at positions $x_n=nd_0$, $n\in{\mathbb Z}$, and their shifts are given by the modulation function $\Theta(x_n)$, where we require that this function is periodic up to a constant shift of the lattice, which means that for every integer $k\in{\mathbb Z}$
\begin{equation}
 \Theta(x+k2d_0/q)=\Theta(x)+k\hat\gamma2d_0/q,
\end{equation}
where $2d_0/q$ is the modulation period, and $\hat\gamma$ is the shift per the period (Figure \ref{fig1bc}(a), $\Theta\equiv{}0$ is set by the reference tetragonal lattice). The value of $\hat\gamma$ represents the slight monoclinic distortion of the 10\,M lattice, as $\hat\gamma$ is the deviation of the respective lattice angle from $\pi/2$, that is, $\hat\gamma=\pi/2-\gamma_{\rm 10M}$). For this reason, we will call it the \emph{monoclinization angle} or simply \emph{monoclinization}.
We can also define {\color{black} shear distortions $\gamma_n$  of individual unit cells  lying along the modulation directions (i.e., modulation function slopes)} as
\begin{equation}
 \gamma_n=\left[\Theta(x_{n+1})-\Theta(x_n)\right]/d_0.
\end{equation}
The monoclinization is then
\begin{equation}
 \hat\gamma=\frac{1}{N}\sum_{n=1}^N{}\gamma_n, \label{eq_hatgamma}
\end{equation}
where we take $N$ equal to $2/q$ for commensurate modulations. 
For incommensurate modulations, the value of $\hat\gamma$ should be approached for $N\rightarrow\infty$. However, as it is shown later, in Ni-Mn-Ga lattice the incommensurateness causes the monoclinization to vary with $x$ {\color{black} (that is, the lattice can be in each point classified as monoclinic, but the monoclinic angle varies in space)}. To capture this behavior,
we will calculate $\hat\gamma(x)$, using (\ref{eq_hatgamma}) in sense of moving averages, with $N$ (assumed to be an odd number, we typically take $N=5$ for modulation {\color{black}wave numbers} close to $q=2/5$) representing the width of the sample window,
\begin{equation}
 \hat\gamma{(x_n)}=\frac{1}{N}\sum_{k=n-(N-1)/2}^{n+(N-1)/2}\gamma_n.
\end{equation}
To visualize the macroscopic shape of the incommensurate lattice  and its dynamic changes, we will than use a cumulative sum of $\hat\gamma{(x_n)}$. This cumulative sum has a meaning of a smoothed approximation of the modulation function itself, showing how the monoclinic angle changes at lengthscales much longer than the modulation period.

\bigskip
In this setting, we consider the periodic part of the modulation function 
\begin{equation}
 \Theta_P(x)=\Theta(x)-\hat{\gamma}x. \label{eq_period_thet}
\end{equation}
{\color{black} This periodic part (typically a sine wave) can be homogeneously shifted in phase with respect to the lattice; as we discuss bellow, specific phase shifts $\varphi$ may represent $a/b$ twins in the commensurate lattice, while in incommensurate lattices $\varphi$ has no physical meaning and can be arbitrary.}
In addition, we consider a small {\color{black} perturbation of the phase shift $\Delta\varphi(x,t)\ll{}\pi{}q$ in $\Theta_P$, representing a phason (see Figure \ref{fig1bc}(b))
\begin{equation}
 \Theta_P(x)\rightarrow\Theta_P(x+\Delta\varphi(x,t)d_0/\pi{}q),
\end{equation}
that may arise in response to external mechanical loading, has a characteristic length, and can propagate through the lattice as a wave.}

 For the two competing interpretations of the modulation discussed above, the behavior of the lattice after such small change of (the periodic part of) the modulation function is very different. For the CDW-based approach, where $\Theta(x)=\Theta_P(x)$ is a sine wave, the phase is free to move back and forth, particularly for the incommensurate lattice where the sine wave cannot pin at lattice sites. The phason is energetically cheap, but does not have any impact on the macroscopic strain of the lattice, since $\hat\gamma=0$ independent of $\varphi$. For the adaptive (nanotwinning) concept, the modulation function is a zig-zag function firmly stuck to the lattice ($\Theta_{\rm NM}$ in Figure \ref{fig1bc}(a)), since rigid planes of the CPS model need to act as twinning planes between NM variants. Any shift of the phase would require large elastic straining of the NM building blocks, that are elastically quite stiff \cite{Sedlak2017}. Since $\hat\gamma\neq{}0$ in the nanotwinning construction, shifting the phase could have effects on the macroscopic strain (i.e., could change $\hat\gamma$), but would be energetically very expensive at the same time. In other words, in none of these concepts, the phasons can effectively relax external mechanical loadings and lead to any particularly soft elastic behavior. To obtain a model in which such a relaxation is possible, we need to introduce coupling between the phase shift and the monoclinization $\hat\gamma$. This is done in the following two sections, merging ideas from the CDW-based interpretation of the modulation function and the adaptive concept. We start first with a perfectly commensurate lattice, construct the static model thereof, then we consider dynamic perturbations of the phase, and finally extend our considerations to incommensurate lattices.

\subsection{Model of the commensurate lattice}

We consider a lattice with $q=2/5$, that means $N=2/q=5$ (a perfect five-layer modulation). We assume that the periodic part of the modulation function is a simple sine wave 
\begin{equation}
\Theta_P(x)=\Theta_0\sin(q\pi{}x/d_0+\varphi), \label{eq_unpert_Theta}
\end{equation}
where $\Theta_0$ is a modulation amplitude, and $\varphi$ is {\color{black} the phase shift of the modulation wave} with respect to the lattice. Physically, the periodic part can be understood as a CDW-induced modulation resulting from a Fermi surface nesting, or from any similar phenomenon. For such choice of $\Theta_P$, we have $\hat\gamma=0$, and the {\color{black} shear distortions} for the individual cells are
\begin{equation}
  \gamma_n=\Theta_0\left[\sin(q\pi{}(n+1)+\varphi)-\sin(q\pi{}n+\varphi)\right]/d_0,
  \label{eq_unpert_gamma}
\end{equation}
being bounded by $|\gamma_n|<\Theta_0q\pi/d_0$ for any $\varphi$, and attaining five different values along the modulation period, which means the unit cells along the modulation wave attain five different shapes, while satisfying
\begin{equation}
 \hat\gamma=\sum_{n=1}^5\gamma_n=0.
\end{equation}

It is natural to assume that some of these shapes can be energetically more favorable then the others. Motivated by the adaptive concept \cite{Kaufman_NJP,Kaufman_PRL}, we will assume that the most favorable cells are those with shapes close to unit cells of the NM martensite (cf. \cite{Gruner}). This means that there are two preferred {\color{black} shear distortions} that we will denote $\pm\gamma_{\rm NM}$, each representing one variant of NM martensite appearing in the nanotwinned material. Following the approaches of the non-linear elasticity theory of martensite, we can consider that there exists a Landau-type energy landscape
\begin{equation}
 E(\gamma)=A(\gamma^2-\gamma_{\rm NM}^2)^2, \label{eq_landscape}
\end{equation}
that has minima for $\gamma=\pm\gamma_{\rm NM}$, and for which the scaling parameter $A$ can be set to adjust for example {\color{black} the height of the energy barrier between the $\pm\gamma_{\rm NM}$ energy wells, as this height is equal to $A\gamma^2_{\rm NM}$}. Because it is known from the experiments \cite{Santamarta,Righi} that the modulation amplitude in 10\,M martensite is smaller than predicted by the adaptive concept, and thus, also the slopes $\gamma_n$ are smaller,  we can choose $\gamma_{\rm NM}>\Theta_0q\pi/d_0$.

\begin{figure}
 \includegraphics[width=\textwidth]{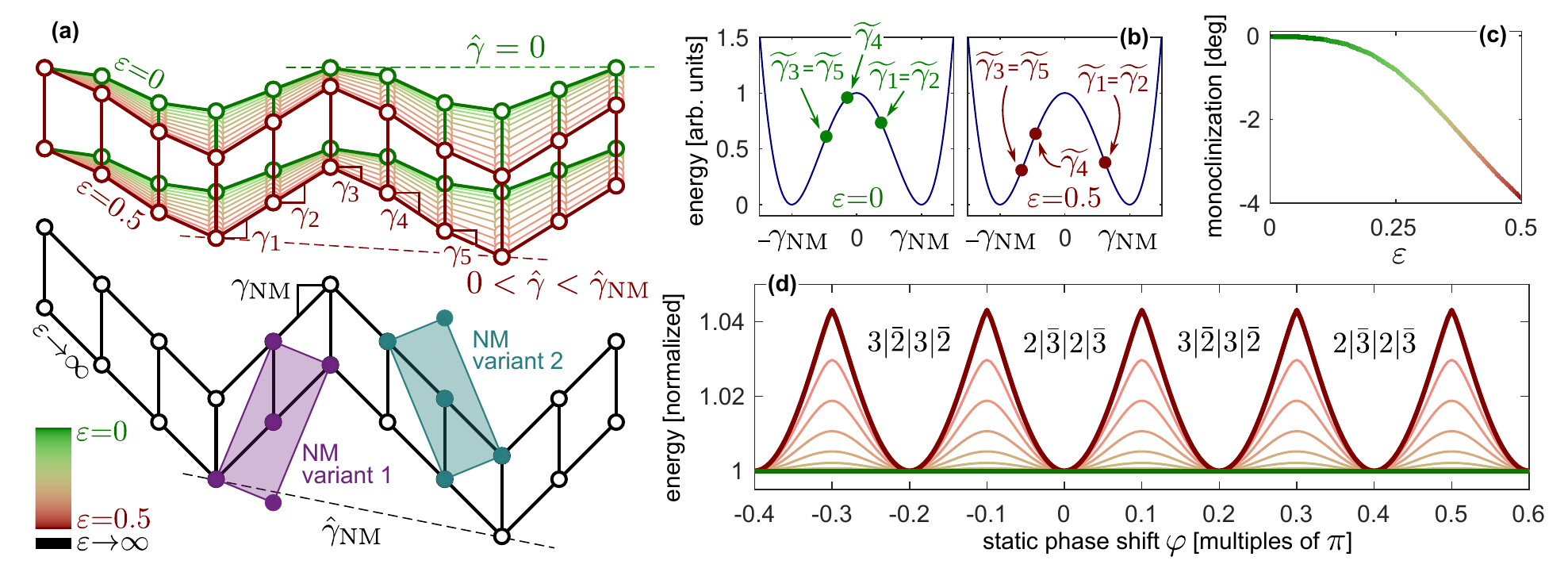}
 \caption{\textbf{(a)} the energy-minimizing modulated lattice for various levels of the weighting factor $\varepsilon$ (for $\varepsilon\rightarrow\infty$ the model collapses to a zig-zag modulation function of the adaptive concept); \textbf{(b)} the energy perturbation landscape, and the optimized values of energy $E_n(\widetilde{\gamma_n})$ for $\varepsilon=0$ and $\varepsilon=0.5$; \textbf{(c)} the macroscopic monoclinization (the monoclinization angle is $\arctan(\hat\gamma)\approx\hat\gamma$) as a funciton of $\varepsilon$; \textbf{(d)} energy $\hat{E}$ as a function of the phase shift $\varphi$ for various levels of $\varepsilon$, the energy is normalized with respect to its value for $\varphi=0$. In \textbf{(a)}, the shapes of the NM martensite unit cells used as building block of the modulated martensite in the $\varepsilon\rightarrow\infty$ limit are drawn. For better visibility of the effect, we use $c_{\rm NM}/a_{\rm NM}>{}2$  (so that $\gamma_{\rm NM}=1$, {\color{black} see Methods for a rationalization}), while the real values are typically close to $c_{\rm NM}/a_{\rm NM}=1.2$  \cite{Kaufman_PRL}. The color scale introduced in the lower left corner applies through subfigures (a)-(d).} \label{fig2}
\end{figure}

\bigskip
As the next step, we will assume that the energy landscape affects the modulation function, that is, that the real modulation function is slightly perturbed by the fact the individual $\gamma_n$ tend lean towards energetically preferred values of $\pm{}\gamma_{\rm NM}$. 
We will denote the perturbed modulation function as $\widetilde{\Theta}(x)={\Theta}(x)+\delta{\Theta}(x)$, and consider resulting {\color{black} shear distortions} $\widetilde\gamma_n=\widetilde{\Theta}(x_{n+1})-\widetilde{\Theta}(x_{n})$. We make the assumption that $\widetilde\gamma_n$ in each cell can be obtained by minimizing energy
\begin{equation}
 E_n(\widetilde\gamma_n)=(\widetilde\gamma_n-\gamma_n)^2+\varepsilon{}A{}(\widetilde\gamma_n^2-\gamma_{\rm NM}^2)^2, \label{eq_balance}
\end{equation}
where the weighting factor $\varepsilon$ is a parameter determining the strength of the perturbation, and $\gamma_n$ are the {\color{black} shear distortions} for the unperturbed (sine-wave) modulation function, calculated \emph{via} (\ref{eq_unpert_gamma}). {\color{black} We assume that the energy perturbations in the individual cells are independent, and thus, the energy-minimizing $\widetilde\gamma_n$ also minimizes the average total energy per unit cell
\begin{equation}
 \hat{E}=\frac{1}{N}\sum_{n=1}^NE_n(\widetilde\gamma_n). \label{eq_energy}
\end{equation}
That means, we assume that the interactions between the cells are fully captured by the unperturbed modulation function, and the additional shear distortions due to the Landau-type energy do not affect them; this is a strong simplification that is justified only for very small perturbations.}

If $\varepsilon$ is sufficiently large, the second term in the energy becomes dominant, and the energy minimizer becomes a close approximation of the zig-zag modulation function predicted by the adaptive concept, as all angles $\widetilde\gamma_n$ fall into the $\pm{}\gamma_{\rm NM}$ energy wells. {\color{black} It is worth noting that the sharp zig-zag morphology with large $\gamma_n$ (close to $\gamma_n=\pm{}\gamma_{\rm NM}$) is also typically predicted for modulated Ni-Mn-Ga martensites by first-principles calculations \cite{Gruner,Kaufman_NJP}; in other words, the first-principles calculations appear to neglect interactions between the cells completely, allowing each cell to reach independently one of the $\pm{}\gamma_{\rm NM}$ energy wells.}

When $1>\varepsilon>0$, the initial modulation function is modified but not overridden by the second term. In this case, the magnitudes of the slopes $\hat\gamma_n$ become increased to lower their misfits from $\pm{}\gamma_{\rm NM}$.  Because $N$ is an odd number, $\widetilde\gamma_n$ that are positive (uphill) can be different in number and in magnitude from those that are negative (downhill), and this leads to asymmetry and to a monoclinic distortion $\hat\gamma\neq{}0$. This distortion is dependent on $\varepsilon$ as well as on other parameters of the model.

For a benchmark test, we set $d_0=1$, $\Theta_0=1/2$, $A=1$, $\varphi=0$, $\gamma_{\rm MN}=1$ {\color{black}(see Methods for rationalization of this setting)}, and observe the evolution of $\hat{\gamma}$ with $\varepsilon$ evolving from 0 to 0.5 (Figure \ref{fig2}(a)): it is seen that for $\varepsilon=0$ the lattice is orthorhombic ($\hat\gamma=0$), while for $\varepsilon=0.5$ the modulation function is already very close to the zigzag function predicted by the adaptive concept, with $\hat\gamma\approx{}4^\circ$, but with a much smaller amplitude of displacements {\color{black}{ than the adaptive concept predicts}}.

In Figure \ref{fig2}(b), the energy perturbation (\ref{eq_landscape}) is shown, and how much are the {\color{black} shear distortions} $\widetilde\gamma_n$ affected by this perturbation with increasing $\varepsilon$. While for $\varepsilon=0$, the points representing $\widetilde\gamma_n$ are close to the top of the energy barrier between the wells, for $\varepsilon=0.5$ they are attracted more towards $\pm\gamma_{\rm NM}$ (and they are exactly at the wells for $\varepsilon\rightarrow\infty$). As seen in Figure \ref{fig2}(a), increasing $\varepsilon$ leads also to a gradual increase of the average monoclinization $\hat\gamma$ (the result is shown in Figure \ref{fig2}(c)). The experimentally observed monoclinic distortions are of the order of 0.2$^\circ$ to 0.4$^\circ$ \cite{Straka2016}, which in our model corresponds to $\varepsilon\approx{}0.2$, and thus, we will {\color{black} futher use this value for demonstrating the main features of the model.}

\bigskip
The individual $\gamma_n$ values are also dependent on the static phase shift $\varphi$, and, so are, consequently, the resulting energy-minimizing $\widetilde\gamma_n$, as well as the averaged quantities $\hat{E}$ and $\hat\gamma$. In Figure \ref{fig2}(d), there is a plot of $\hat{E}$ with respect to $\varphi$ for the same range of $\varepsilon$. It is seen that for any $\varepsilon>0$ the modulation wave can get pinned at specific phase shifts with respect to the lattice. The energy minima appear at $\varphi=0$ (sine wave starting at one of the rigid lattice planes) and at $\varphi=k\pi/q=2k\pi/5$, which represents lattice-invariant shifting by an integer multiple of lattice spacings $d_0$. This justifies our choice of $\varphi=0$ for initial calculation of the modulation function and monoclinization angles shown in Figures \ref{fig2}(a,c): $\varphi=0$ is one of the phase shifts that the lattice tends to achieve to minimize its energy.

In addition to $\varphi=k\pi{}q=2k\pi/5$, there are equivalently deep minima also for  $\varphi=(2k+1)\pi/5$; it can be easily understood that for $\varepsilon\gtrsim{}0.5$ these minima correspond to mirror
reflection of the nanotwin stacking sequence from $2|\bar{3}|2|\bar{3}$ to $3|\bar{2}|3|\bar{2}$. For general $\varepsilon$, the $\varphi=\pi/5$ static phase shift represents $a/b$ twinning, which means the sets of minima at $\varphi=2k\pi/5$ and $\varphi=(2k+1)\pi/5$ represents two variants of 10\,M martensite that differ in $a$ and $b$ lattice parameter orientations, but keep the same orientation of the modulation vector. The lattices of these variants are mirror-reflected about the plane perpendicular to the modulation vector, and have the same magnitude but opposite signs of the monoclinic angle $\hat\gamma$. It is also worth noting that the height of the energy barriers between the minima steeply increases with increasing $\varepsilon$. This means that while the sharp nanotwins (large $\varepsilon$) considered in the adaptive concept are strongly pinned to the lattice, those with smoother modulations can more easily jump between neighboring minima, which agrees well with the experimentally observed high mobility of $a/b$ twins \cite{Saren}. 

We can conclude that we have constructed a model of a commensurate 10\,M lattice that captures all its main topological features: for proper choice of $\varepsilon$, it has energy minimizers in forms of modulated lattices that resemble $2|\bar{3}|2|\bar{3}$ and $3|\bar{2}|3|\bar{2}$ nanotwin stacking sequences of the adaptive concept, but are smoother and have smaller amplitudes, which is in agreement with experimental observations \cite{Santamarta,Righi,Vertat2024}. These minimizers are macroscopically slightly monoclinic, with monoclinic angles close to those observed experimentally.  Importantly, these properties of the energy minimizers were not \emph{a priori} postulated in the model construction, they spontaneously arise as results of the energy balance between the first and the second term in (\ref{eq_energy}). In other words, the model takes a harmonic modulation resulting from the CDW-based concepts and let it interact with the energy landscape considered for NM martensite.  

\subsection{Phason mechanics and extension to incommensurate lattices}

Let us now assume a low-amplitude phason $\Delta\varphi(x,t)$ acting on the unperturbed modulation function (\ref{eq_unpert_Theta}), that means a small phase shift in the CDW-related, sine-like modulation wave. If the phason amplitude is much smaller than $\pi/10$, {\color{black} the energy needed to propagate the phason is much smaller than to propagate an $a/b$ twin interface. At and above $\Delta\varphi=\pi/10$, in contrast, the maximum of energy in Figure (\ref{fig2}(d)) is reached, and the system can spontaneously evolve into the neighboring minimum, which means changing the stacking sequence, i.e., the $a/b$ twin interface moves.}

\begin{figure}
 \centering
 \includegraphics[width=\textwidth]{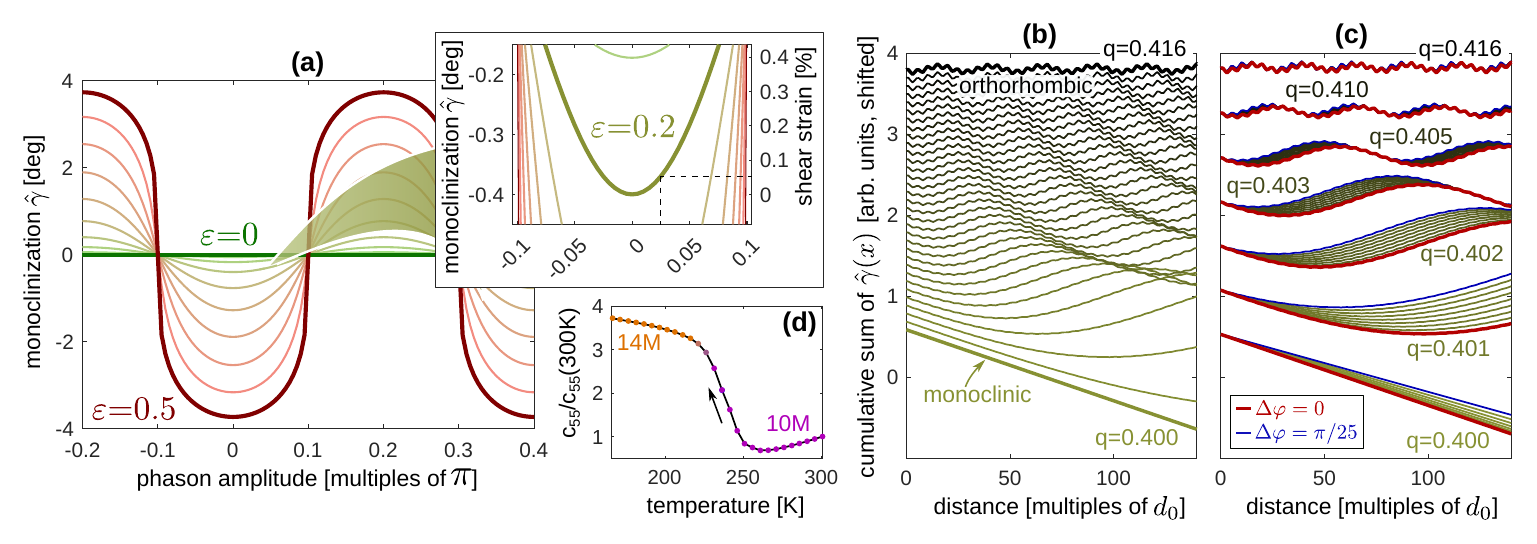}
 \caption{\textbf{(a)} the lattice monoclinization changes under a small-amplitude phason (the magnified plot in the upper right corner helps resolving the details for $\varepsilon=0.2$, the right vertical axis recalculates the monoclinization to the relative shear strain); \textbf{(b)} the effect of incommensurateness on the macroscopic shape of the lattice, from perfectly commensurate ($q=0.400$) to the incommensurate state discussed in \cite{Sozinov2026} ($q=0.416$); \textbf{(c)} change of the macroscopic shape of the lattice under a small-amplitude phason (individual lines represent phasons with amplitudes between zero and the maximum with a $\pi/200$ step); \textbf{(d)} temperature evolution of the softest shear elastic coefficient $c_{55}$ in Ni-Mn-Ga-Fe presented in \cite{Nejezchlebova_ESOMAT}, with cooling from the 10\,M commensurate phase, over incommensurate, to 14\,M. The value of $c_{55}$ is normalized with respect to $c_{55}=2.2$ GPa at 300~K.}\label{fig3}
\end{figure}

However, even the phasons with such small amplitudes can carry macroscopically relevant changes in $\hat\gamma$. This is proved by a direct calculation summarized in Figure \ref{fig3}(a), which shows the monoclinization $\hat\gamma$ of a commensurate lattice under small phase shifts (phason amplitudes). If the phase shift amplitude reaches over $\pm\pi/10$, the monoclinization reverts form $\hat\gamma<0$ to $\hat\gamma>0$, which is an $a/b$ twin reorientation. Nevertheless, for the selected $\varepsilon=0.2$ the phason with amplitude $\pi/40$ carries shear strain of $5.3\times{}10^{-4}$ (see the magnified area in Figure \ref{fig3}(a)), which is fully comparable to maximum strain amplitudes used in mechanical dynamic tests \cite{Kustov}, and two orders of magnitude higher than those used in ultrasonic \cite{Dai} and laser-ultrasonic \cite{Repcek2024} characterizations. This means that these strains in the experiments can be fully or partially relaxed by low-amplitude phasons, and the effective elastic behavior may become very soft. 

In other words, while the shearing of the individual unit cells along the modulation planes can be quite stiff, as proved by INS experiments and by first-principles calculations, their collective motion due to small phase shifts in the modulation function can be energetically very cheap. Our model suggest that this is possible because of an interplay between the CDW-induced harmonic modulation of the lattice and the existence of the energetically favorable ground state, which is the NM martensite. If the modulation was symmetric with respect to the lattice, having periodicity equal to an even number of interatomic spacings such as the 4O structure observed in Ni-Mn-Sb martensites \cite{NiMnSb4O}, this feature would not arise, because the effect of the  $\pm\gamma_{\rm NM}$ minima would be symmetric, not leading to any macroscopic monoclinic distortion $\hat\gamma$. Similarly, in materials with odd-number-layered structure, but without any higher-symmetry ground state the existence of which perturbs the modulation function, there is no spontaneous monoclinic distortion due to the modulation, and, consequently, the phasons cannot relax the externally imposed macroscopic strains. These consideration suggest that the 10\,M martensite in Ni-Mn-Ga (and Ni-Mn-Ga-based alloys) possesses a unique combination of prerequisites needed for phason-mediated soft mechanics, probably not met by any other metallic material known.

\bigskip
{\color{black} The key condition for the above described behavior is the high mobility of the phasons, that is, very weak pinning of the phasons to the lattice. Phasons can be considered as completely free to move in lattices with incommensurate modulations, where there is no energetically preferred phase difference between the lattice and the modulation wave. However, the presence of gap-less phason-related branches in INS results on close to pseudo-commensurate ($q=3/7$) and commensurate crystals \cite{Shapiro,Ener2013} indicate that the pinning is very weak even for modulation waves that match small numbers of lattice spacings. On contrary, the soft elastic behavior disappears when the modulations become incommensurate ($q=0.416$) \cite{Sozinov2026}, indicating that releasing the phasons from the pinning due to incommensurateness is not the main mechanism behind the observed features.}

In Figure \ref{fig3}(b), we consider $\varepsilon=0.2$, {\color{black}constant phase ($\varphi=0$, $\Delta\varphi=0$),} and a modulation {\color{black}wave number} evolving from $q=2/5$ to $q=0.416$ (close to the elastically stiff material reported in \cite{Sozinov2026}). For each value of $q$ the monoclinic distortion is calculated in a sense of moving averages with $N=5$, as explained below Eq. (\ref{eq_hatgamma}). Because the monoclinizations themselves are very small (due to our physically relevant choice of $\varepsilon$), we do not plot $\hat\gamma(x)$ directly, but its cumulative sum over more than one hundred $d_0$. This allows us to see the macroscopic character of the lattice and its long-distance evolutions. It is seen that for $q=2/5$ the lattice is homogeneously monoclinically distorted, as the cumulative sum of $\hat\gamma(x)$ smoothly and constantly grows. Then, with even very small deviation from commensurateness (the second plotted curve corresponds to $q=0.4005$) the monoclinic distortion becomes heterogeneous. When approaching $q=0.416$, the lengthscales at which the lattice is monoclinic scales down to $\sim{}10d_0$, which makes the material macroscopically orthorhombic. Moreover, with increasing $q$, the cumulative sum of $\hat\gamma(x)$ becomes oscillatory, that is, the monoclinization angle $\hat\gamma$ becomes strongly dependent on which five unit cells are taken to calculate it in (\ref{eq_hatgamma}).

In other words, for $|q-2/5|\ll{}1$, there still exist extensive regions with positive or negative $\hat\gamma(x)$, and in these regions the monoclinic angle can be relaxed through a collective shift of the modulation phase, that is, by a phason, similarly as for the commensurate phase. For $q$ approaching 0.416, such regions disappear, and in turn, although the phasons in such a system can most probably move quite freely through the lattice (as suggested also from the INS experiments \cite{Shapiro}), they do not carry any macroscopic strains, and the lattice is macroscopically stiff. At the atomisitic scale, the lattice is compliant to small atomic shuffles carried by the phasons that cannot, however, relax external mechanical loadings.

A more detailed insight into this behavior is shown in Figure \ref{fig3}(c), where we show the change of the cumulative sum of $\hat\gamma(x)$ under a small ($0\leq\Delta\varphi\leq\pi/25$) phase shift in $\Theta_P(x)$. It is seen that the lattice macroscopically moves (at distances given by the length of the plotted domain) for $q<0.405$; for higher values of $q$, the changes due to the phason become comparable to the amplitude of the waviness of the cumulative sum itself, and vanishes to zero. 

For $q$ between perfectly commensurate and strongly incommensurate, the length of the phason $L_{\Delta\varphi}$ becomes an important parameter. If we consider the phason to be infinitely broad, spanning over the whole crystal, the cumulative sum of $\hat\gamma(x)$ averages out to zero even for $q-2/5\ll{}1$, and relaxation of macroscopic strains becomes impossible for any other lattice than perfectly commensurate. In real crystals, however, the length is limited by the coherent length of the lattice, which might be a distance between neighboring defects, such as antiphase boundaries or composition variations, all falling to the range of hundreds of nanometers or few micrometers, that is, hundreds to thousands of $d_0$. On the other hand, the phason needs to be a collective motion of the phase over numerous periods of the modulation function, which means the shortest meaningful $L_{\Delta\varphi}$ is at about $10Nd_0=50{}d_0$. 

In the modelled situation in Figure \ref{fig3}(c), the phasons of length $L_{\Delta\varphi}\approx{}100{}d_0$ (that is, long enough for a collective behavior but shorter than distances between defects) induce the most pronounced motion of the lattice for a slightly incommensurate lattice ($q=0.401$). Consequently, it seems like the most effective relaxation of external strains through phasons may appear for this level of incommensuratenes, not for exactly $q=2/5$. This might be related to the experimentally observed behavior of $c_{55}$ with in 1\%Fe-alloyed Ni-Mn-Ga presented in \cite{Nejezchlebova_ESOMAT} and shown in Figure \ref{fig3}(d) (see also \cite{Vertat} for a similar result). With cooling, starting from perfectly commensurate modulations and evolving towards pseudo-commensurate $q=3/7$, the $c_{55}$ first slightly decreases, and at about 260 K it starts steeply increasing, and increases approximately three times until the transition temperature for the intermartensitic transition from 10\,M to 14\,M is reached, where the slope becomes more gradual. There is no direct proof that the initial softening of $c_{55}$ with cooling is caused by slight incommensurateness and by the effect seen in Figure \ref{fig3}(c). However, it can be understood as another point where the proposed simple model does not contradict the experimental data.

The above considerations also suggest that the characteristic coherent lengths of the phasons $L_{\Delta\varphi}$ may play an important role in the macroscopic elastic stiffness of the lattice for fixed $q$. $L_{\Delta\varphi}$ is bounded from above by the distance between defects; if this distance varies through the crystal, or if it is changed for example by surface treatment or by room-temperature relaxation of the surface after the treatment, the local laser-ultrasonic experiments, such as the transient grating spectroscopy measurements applied in \cite{Repcek2024}, may give different results at different locations or in different times. That can rationalize the heterogeneity in the elastic behavior reported in \cite{Repcek2024}.

\bigskip
Regardless of the initial softening and the heterogeneity, the main effect of the transition from $q=2/5$ towards $q\rightarrow{}3/7$ is the stiffening, which we above interpret as a result of loss of the monoclinization of the lattice. Indeed, for phason lengths not smaller than $10Nd_0$ and for $q\geq{0.410}$, the phason necessarily reaches over several 'uphill' and 'downhill' segments of the lattice (see in the cumulative sum of $\hat\gamma(x)$ shown in Figure \ref{fig3}(b)), and the phase shift carried by the phason does not affect the macroscale monoclinization of the lattice which remains equal to zero. 

It is worth noting, however, that the stiffening due to the commensurate$\rightarrow$incommensurate transition reported in \cite{Vertat,Nejezchlebova_ESOMAT,Sozinov2026} is still not as large as the difference between the experimental results for the commensurate structure from ultrasonic measurements from \cite{Repcek2024} and $c_{55}\geq{}15$\,GPa reported from DFT calculations and INS experiments. This means that there still exists an inelastic straining mechanism causing strong elastic softening even in the incommensurate lattice. A possible explanation is that the incommensurate (or pseudo-commensurate) structure is asymmetrically faulted, maintaining, thus some macroscopic monoclinic distortion, although the lattice parameters determined from x-ray diffraction are perfectly orthorhombic; this might be because the incommensurate lattice forms gradually from the commensurate one, and thus, it may adopt part of its asymmetry. The propagating phasons can then interact with these faults, leading to much weaker but still observable elastic softening. Such considerations, however, fall beyond the scope of this paper and beyond the capabilities of the proposed simple model.

\section{Discussion and concluding remarks}

Above, we have hypothesized about the possible contribution of modulation phasons on the soft elastic behavior of 10\,M Ni-Mn-Ga martensite. The proposed model shows that such a contribution is possible, and is perfectly in-line with experimentally observed properties of the 10\,M lattice: its slight monoclinic distortion in the commensurate state, the spontaneous creation of $2|\bar{3}|2|\bar{3}$ and $3|\bar{2}|3|\bar{2}$ sequences, and stiffening and orthorhombization of the lattice with the departure of $q$ from exactly $q=2/5$, as observed upon cooling.

There are, however, two features that the model cannot capture and that challenge the proposed concept of mechanical phason. They are the effect of $a/b$ twins on the macroscopic elastic behavior, and the contribution of the proposed phason mechanics to the high mobility of the (supermobile) $a/c$ twins. Regarding the former, this question has been mainly discussed by Sozinov et al.\cite{Sozinov2026}, who proposed that the very low elastic constant  $c_{55}$ reported in \cite{Repcek2024}, as well as the stiffening with cooling reported by Ve\v{r}t\'{a}t et al.\cite{Vertat}, may be, in fact, partially caused by reversible motion of $a/b$ twins, as these twins carry the same orientation of shear strain as the modulation itself. With incommensurate modulations, when the lattice becomes orthorhombic, the shear strain carried by the motion of an $a/b-$twin vanishes to zero, and thus the contribution of the $a/b$ twins to the soft elasticity is suppressed. Indeed, when the $a/b-$twins were removed from the commensurate 10\,M structure by a small unidirectional pre-strain (favoring the $a-$orientation over the $b-$orientation along the pre-strain direction, see the Supplementary information to \cite{Repcek2024}), some stiffening of $c_{55}$ was observed, and the resulting values of $c_{55}$ were comparable to those of incommensurate martensite. Moreover, the ability of the $a/b$ twins to move under dynamic mechanical loading is well known \cite{Kustov}, and so is the very low twinning stress needed to move these boundaries \cite{Saren}.  

In opposition to this explanation, Straka et al.\cite{Straka2018} and Ve\v{r}t\'{a}t et al.\cite{Vertat} observed a gradual but significant increase of the density of $a/b-$twin interfaces with the $q\rightarrow{3/7}$ evolution upon cooling, when the width of the twin domains can go down to nanoscale. Firstly, this means that in strongly incommensurate structures (as the one with $q\approx{}5/12$ studied in \cite{Sozinov2026}), the $a/b$ twin domain width corresponds just to ten or less 10\,M unit cells, each being differently shifted in phase with respect to the modulation wave, that is, each having slightly different shape. Consequently, the lattice parameters $a$ and $b$ are difficult to define, and the lattice is orthorhombic with $a=b$ only at the averaged, mesoscopic scale, not being relevant to the strains carried by atomisitic-scale motion of the twins. Also, from the opposite side, Straka et al.\cite{Straka2018} show that the number of $a/b$ twin interfaces in perfectly commensurate lattices can be very small, and so it is difficult to ascribe the particular shear softness of this lattice to their contribution. 

Secondly, the observation in \cite{Straka2018,Vertat} means that the $a/b$ twins are directly related to the incommensurate modulations. Possibly,  as discussed in \cite{Straka2026}, they are both just consequences of one phenomenon. Indeed, for $q>2/5$, the phase shift along the modulation wave accumulates, and after sufficiently long distance it reaches values close to $\pi/10$, which means the modulation sequence is mirrored, and the $a-$domain turns into a $b-$domain, which corresponds to the 'uphill' and 'downhill' segments in Figure \ref{fig3}(b). In other words, the width of $a/b$ domains in incommensurate modulations is necessarily limited, and it goes to zero with increasing incommensurateness. As a result, it is not fully justified to consider $a/b$ twins as defects that are present in the lattice and are mobile in the commensurate phase while immobile in the incommensurate phase, where the driving force diminishes due to $a\rightarrow{}b$, as considered in \cite{Sozinov2026}. 

From the point of view of the proposed model of mechanical phasons, however, the $a/b$ twins necessarily strongly affect the elastic response of the 10\,M lattice. In the commensurate state, where the $a/b$ domains are broad, and phasons with large $L_{\Delta\varphi}$ can equivalently relax strains in both of $a-$oriented and $b-$oriented domains. As $\hat\gamma<0$ in the $a-$domains and $\hat\gamma>0$ in the $b-$domains, the presence of $a/b$ twins helps the macroscopic response to be symmetric, linear, and soft regardless of the orientation of shear loading. When the $a/b$ twins are driven away by pre-strain, as in the Supplementary material to \cite{Repcek2024}, only one half of the possible relaxation mechanism can be utilized (only specific sign of shear strain carried by the elastic wave can be relaxed), which could lead to the observed stiffening of $c_{55}$.

\bigskip
Regarding the relation between the phason mechanics and highly mobile $a/c$ interfaces, the situation is much less clear. It is known that the high mobility  is not affected by the commensurateness of the modulation wave, at least for the Type-II twins \cite{Straka2016}. The Type-II twin interfaces are quite generally oriented with respect to the 10\,M lattice of the matrix, and thus, it is difficult to unravel how the shifts in the modulation wave running along a single crystallographic direction on each side of the interface could facilitate its motion. The mechanism of the $a\rightarrow{c}$ reorientation proposed in \cite{Repcek2024} works only for the considered discrete faulting of the modulation sequence, not for a collective motion of the phase of the modulation wave. 
For Type-I twins, for which there can exist pinning of the $a/c$ interface on the $a/b$ twins \cite{Seiner2014}, it is clear that the rigidity of $a/b$ interfaces (and thus, the strength of the pinning) is affected by the elastic constants of the single monoclinic variants, and by $c_{55}$ in particular. Most probably, however, there is a more delicate connection between the phasons and the supermobility than just through the elastic stiffness of the martensitic variants. The elastic softness in $c_{55}$ is unique to Ni-Mn-Ga, and so is the supermobility of the twin interfaces. If the former is related to phasons, and the model introduced in this paper indicates it can be, a strong motivation arises also to discuss how the shifts of the phase of the modulation wave can enhance the mobility of the $a/c$ twins. In this paper, we have presented a tool to model lattice mechanics related to phasons; a logical next step would be to implement this model into three-dimensional atomistic-scale simulations of the motion of $a/c$ twin interfaces. The fact that the supermobility is preserved in strongly incommensurate lattices, where $c_{55}$ is stiff and phasons cannot relax macroscopic shear strains, suggest that the energetically cheap mechanism lowering the pinnig between the lattice and the $a/c$ twin might be the phason-related shuffle of the atoms. 

\section{Summary}

The simple theoretical model introduced in this paper aimed at explaining the soft elastic behavior in 10\,M martensite of Ni-Mn-Ga through phase shifts of the modulation function, which means, through modulation phasons. We started with constructing a model of a commensurate 10\,M lattice ($q=2/5$), making three main assumptions:
\begin{enumerate}
 \item the crystal spontaneously tends to form periodic modulations (be it because of the CDW or any other reason), with a low-amplitude sine wave (\ref{eq_unpert_Theta}) as the modulation function;
 \item the static modulation wave interacts with the energy landscape so that some shapes of the unit cells (some $\gamma_n$) are energetically preferred over others; in particular we assumed that the energetically preferred unit cells are those with shapes closest to the equivalent cells of NM martensite, as introduced by the adaptive concept;
 \item the resulting (perturbed) modulation function $\widetilde\Theta{}(x)$ can be obtained by energy minimization, balancing the energy contributions from harmonic modulations and from the energy landscape through equation (\ref{eq_balance}).
\end{enumerate}
We calculated the function $\widetilde\Theta{}(x)$ for the weighting factor ranging from $\varepsilon=0$ (zero perturbation) to $\varepsilon=0.5$ (strong perturbation from the energy landscape). We observed its properties, as summarized in Figure \ref{fig2}. The main observations were:
\begin{enumerate}
\setcounter{enumi}{3}
\item the modulated lattice becomes spontaneously monoclinic (with experimentally realistic monoclinic angles obtained for $\varepsilon\approx{}0.2$), it is easily seen that this because $N=2/q$ is a low odd number;
\item with increasing $\varepsilon$, the modulation function quickly evolves towards a sharp zig-zag function, similar to that predicted by the adaptive concept, but with lower amplitude of modulations and lower resulting monoclinization angle $\hat\gamma$; this agrees well with experimental observations;
\item for considered static, constant phase shifts $\varphi$, the lattice attains minimal energy for $\varphi=2k\pi/5$ (with $\hat\gamma<0$) and $\varphi=(2k+1)\pi/5$ (with $\hat\gamma>0$), which corresponds to $a-$orientations and $b-$orientations in $a/b$ twins, respectively; for physically relevant levels of $\varepsilon$, the barrier between minima is much smaller than in the $\varepsilon\rightarrow\infty$ limit of the adaptive concept, which means that the twinning stress for the $a/b$ reorientation is much smaller than the twinning stress in NM martensite, which is also observed experimentally.
\end{enumerate}
Finally, we discussed the effect of small, localized, {\color{black}perturbations of the} phase shift $\Delta\varphi(x,t)$ representing phasons. We considered phasons of coherent lengths $L_{\Delta\varphi}$ over at least ten spatial periods of the modulation wave, and shorter than typical distances between defects (hundreds or thousands of spatial periods). For such phasons, the model predicts that:
\begin{enumerate}
\setcounter{enumi}{6}
 \item even very small phason amplitudes ($\varphi=\pi/40$) carry monoclinic strains of order $10^{-4}$ (Figure \ref{fig3}(a)), which means they can fully relax strains in laser-ultrasonic experiments; indeed, as the experimentally measured shear modulus $c_{55}\approx{2}$\,GPa \cite{Dai,Repcek2024} is ten times softer than $c_{55}\approx{20}$\,GPa obtained from first-principles calculations \cite{OzdemirKart,SeinerReview} or neutron-scattering experiments \cite{Shapiro,Ener2013},  we search for mechanism able to relax up to 90 \% of strains inelastically;
 \item once the modulations become incommensurate, the macroscopic monoclinic distortion gradually disappears, and so does the ability of the phasons to relax the strains (Figure \ref{fig3}(b,c)); this agrees with the experimentally observed stiffening with the commensurate$\rightarrow$incommensurate transition.
\end{enumerate}
It is important to point out that both the {\color{black} static phase shift $\varphi$ and its dynamic perturbation $\Delta{}\varphi$} in our model were considered in the unperturbed (sine-wave) modulation function $\Theta_P(x)$, while the perturbations $\delta\Theta(x)$  in response to them, including the monoclinic distortion, resulted solely from the interaction of $\Theta_P(x)$ with the energy landscape. This means that we can construct the whole mechanics of the five-layer modulated lattice just from assuming a sine-wave modulation function with very weak pinning between the phase and the lattice. Then, small dynamic phase shifts in $\Theta_P(x)$ can relax shear strains corresponding to $c_{55}$ in elastodynamic experiments, while large static phase shifts in $\Theta_P(x)$ represent $a/b$ twinning -- this makes the model particularly simple: it is assumed that the lattice has just one additional degree of freedom, which is the phase of the modulation wave, and all macroscopically observed features are then just the consequences of the phase shifts. The model is in agreement with the current theoretical considerations of equivalency between incommensurateness and density of $a/b$ boundaries \cite{Straka2026}, as well as with seemingly contradictory experimental results obtained for commensurate  and incommensurate 10\,M structures \cite{Repcek2024,Sozinov2026}. However, it remains a challenge to apply the conclusions drawn from this model to interpret the supermobility of the $a/c$ twins in Ni-Mn-Ga; our preliminary considerations indicate that it might be the phason-mediated shuffling of atoms behind this unique phenomenon.

\section{Methods}
\subsection{Numerical simulations and the choice of parameters}
The coupling between mechanical shears and modulation wave discussed in this paper is presented through a model based on minimization of the energy (\ref{eq_balance}), where the differences $\gamma_n$ are given by the unperturbed modulation function (\ref{eq_unpert_Theta}). As the energy is a simple fourth-order polynomial, the problem is partially analytically treatable and reduces to finding roots of the derivative
\begin{equation}
 {\rm d}E_n/{\rm d}{\widetilde{\gamma}_n}=4\varepsilon{}A{\widetilde{\gamma}_n}^3 +2\left(1-2\varepsilon{}A\gamma_{n}\right)\widetilde{\gamma}_n -2\gamma_{n}, \label{eq_derivative}
\end{equation}
which is a depressed cubic equation, but, in general, does not have analytical solutions in the sense of Carnado's formula and needs to be solved numerically, and afterwards, the root representing a minimum closest to the given $\gamma_n$ needs to be identified.

{\color{black}{As $\varepsilon$ and $A$ appear only in a product, we can always choose $\varepsilon$ to represent a small perturbation running from 0 to 0.5, and then the energy has three adjustable parameters: the amplitude of the unperturbed modulation function $\Theta_0$ (that affects the energy through $\gamma_n$), the slope $\gamma_{\rm NM}$ representing nano-twinning of non-modulated martensite, and the scaling parameter $A$, that determines the height of the barrier between the $\pm\gamma_{\rm NM}$ energy wells.

According to \cite{Vertat2024}, the amplitude of the modulation function at the room temperature is approximately 10 \% of the interatomic spacing, which implies the choice of $\Theta_0 = 0.1{}d_0$. The value of $\gamma_{\rm NM}$ can be deduced from the $c/a$ ratio of non-modulated martensite; this ratio is typically close to 1.2, which is achieved for $\gamma_{\rm NM}=0.18$  \cite{Niemann_Jalcom}. Regarding the parameter $A$, the height of the barrier is unknown from the experiments, and thus, $A$ can be used as a fitting parameter for the outputs of the model. For example, the monoclinic angle in the commensurate phase reported from experiments deviates from $\pi/2$ by less than 1$^\circ$, and thus, a justified choice if $A$ is such that the range $0\leq{}\varepsilon\leq{}0.5$ convers the range of $0\leq{}\hat{\gamma}\leq{}1^\circ$. This is approximately achieved for $A=40$.

The results of the model (with zero phase shift, $\varphi=0$) are shown in Figure \ref{fig_methods}. As expected, with increasing $\varepsilon$ the character of the modulation function changes from sine-wave to zig-zag, as the energy minimizing $\widetilde{\gamma}_n$ are pushed towards the
$\pm\gamma_{\rm NM}$ wells (Figure \ref{fig_methods}(b)). The resulting monoclinization $\hat{\gamma}$ evolves from 0 to 1$^\circ$, and reaches experimentally most relevant values (between 0.2$^\circ$ and 0.4$^\circ$) close to $\varepsilon=0.2$.

\begin{figure}[!b]
\begin{minipage}[b]{0.6\textwidth}
 \hspace{-0.7cm} \includegraphics[width=1.0\textwidth]{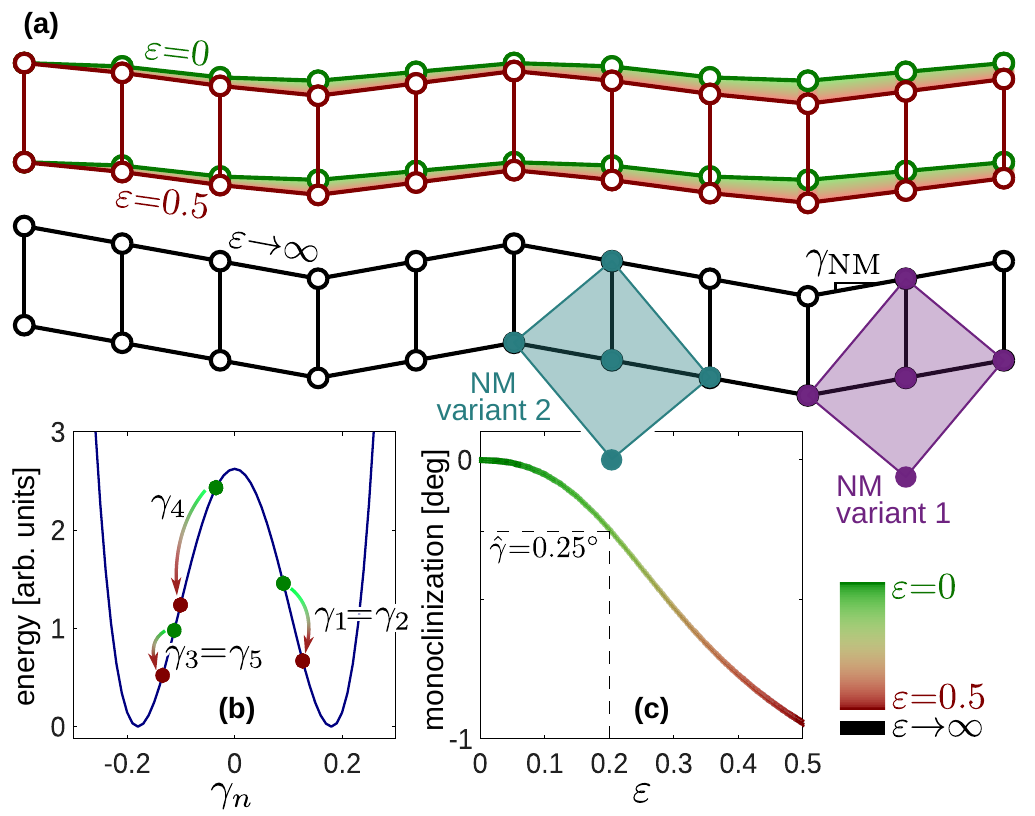}
 \end{minipage}\hfill
\begin{minipage}[b]{0.4\textwidth}
  \caption{{\color{black}{Behavior of the numerical model for parameters giving the best approximation of the experiment in terms of lattice parameters, using the same visualization as in Figure \ref{fig2}: \textbf{(a)} evolution of the modulated lattice from sine-wave modulations ($\varepsilon=0$) to nearly zig-zag ($\varepsilon=0.5$), and comparison to the nanotwinning geometry; \textbf{(b)} energy-minimizing distortions $\gamma_n$ of the individual unit cells, and their evolution on the Landau energy landscape with increasing $\varepsilon$; \textbf{(c)} resulting monoclinization $\hat{\gamma}$, reaching the experimentally observed magnitude for $\varepsilon=0.2$.}}}\label{fig_methods}
  \end{minipage}
\end{figure}

At the same time, however, it is seen that this set of parameters is not optimal for demonstrating visually the behavior of the model, as both the slopes $\gamma_n$ and the monoclinic distortions $\hat\gamma$ are very close to zero. The sine-wave character of the modulation function at $\varepsilon=0$ is not clearly visible, and nor is the difference between the zig-zags for $\varepsilon=0.5$ and $\varepsilon\rightarrow\infty$. In turn, this setting is unsuitable for visualizations of for example the spatial evolution of $\hat{\gamma}(x)$ in incommensurate structures, or for discussing of evolutions of several calculated parameters with $\varepsilon$, as done in the main text. For this reason, we decided to use a different setting for the benchmark calculations; the setting is required not to change the balance between the harmonic (CDW based) term and the Landau energy term, to give similar monoclinic distortion $\hat{\gamma}$ for $\varepsilon\approx{0.2}$, but to predict much larger slopes $\gamma_n$. Such a setting can be obtained by increasing equivalently $\Theta_0$ and $\gamma_{\rm NM}$, while decreasing (quadratically) $A$: this is approximately satisfied with a simple set of $\Theta_0/d_0=1/2$, $A=1$ and $\gamma_{\rm NM}=1$ used in the main text. As seen in Figure \ref{fig2}, with this setting the model behaves equivalently as for the initial setting (in qualitative terms), but the changes in the slopes as well as in the character of the modulation function are much clearer. For this reason, we use the latter setting for the simulations reported in the main text; it can be, nevertheless, easily checked, that all important features the model predicts for the 'benchmark setting' in the main text can be also predicted for the physically relevant setting introduced above.}}

\subsection{Experimental methods}

In this paper, we refer to several experimental results that were obtained in Ni-Mn-Ga-based single crystals; for the analysis of the modulation {\color{black}wave number} and modulation function, the diffraction experiments are in detail described in \cite{Vertat,Straka2026,Vertat2024}; for the laser-ultrasonic characterization of soft elastic modes in commensurate 10\,M martensite, details were reported in \cite{Repcek2024}.

For the temperature dependence of $c_{55}$ along the intermartensitic 10\,M $\rightarrow$ 14\,M transition (shown in Figure \ref{fig3}(d) and presented in \cite{Nejezchlebova_ESOMAT}), the experimental details are not provided in available literature, and thus, we summarize them here briefly: the temperature dependence was obtained using laser-based contact-less resonant ultrasound spectroscopy, performed in a cryo-chamber cooled by nitrogen vapours. The sample was an approximately 3$\times$2$\times$1 mm$^3$ single crystal, and $c-$axis was set perpendicular to the largest face of the sample at the room temperature (that is, in the 10\,M structure). The sample was freely laid in the chamber, so that the intermartensitic transition was not affected by external stresses. The cooling was performed with average rate of 1 K.min$^-1$, for which the heat transfer between the sample and the cooling stage was ensured by a low-pressure (20 mbar) nitrogen atmosphere. Infrared (pulsed) laser beam was used for generating the vibrations, and red, continuous laser beam for detection. The evolution of  $c_{55}$ was extracted from the evolutions of the lowest resonant frequencies that are, for strongly anisotropic materials, fully determined by the softest shear elastic coefficient (see also \cite{Repcek2024}).

\subsection*{Acknowledgement}
This work has been financially supported by Czech Science Foundation [project No. 24-10334S] and by Czech Ministry of Education, Youth and Sports through project FerrMion [No. CZ.02.01.01/00/22\_008/0004591] co-funded by European Union.

 \subsection*{Declarations} 

{\bfseries{Funding }} --  This work has been financially supported by Czech Science Foundation and by Czech Ministry of Education, Youth and Sports, through projects specified in the above Acknowledgement .

\medskip
\noindent{\bfseries{Conflicts of interest/Competing interests }} --  The authors declare that they have no conflict of interest.

\medskip
\noindent{\bfseries{Data availability}} --  Not applicable (paper does not include original research data).  

\medskip
\noindent{\bfseries{Code availability}} --  Codes used for visualizations of the model behavior shown in Figures \ref{fig2} and \ref{fig3} can be shared upon reasonable request.

\medskip
\noindent{\bfseries Authors Contributions} -- All authors contributed equally to the study. The theoretical concept was introduced by Petr Sedl\'{a}k, the micromechanical model and the mathematical modelling framework were developed by Hanu\v{s} Seiner, and the discussion of the results with respect to experiments was carried out by Tom\'{a}\v{s} Grabec. The first draft of the manuscript was written by Hanu\v{s} Seiner and all authors commented on previous versions of the manuscript. All authors read and approved the final manuscript.

\end{document}